\documentclass[10pt]{article}

\usepackage{amsmath}
\usepackage{amssymb}
\usepackage{mathrsfs}
\usepackage{graphics}
\usepackage{tensor}
\usepackage{amsbsy} 
\usepackage{mathrsfs} 

\begin{document}

\title{\textbf{Bell inequalities with retarded settings}}
\author{Lucien Hardy\\
\textit{Perimeter Institute,}\\
\textit{31 Caroline Street North,}\\
\textit{Waterloo, Ontario N2L 2Y5, Canada}}
\date{}

\maketitle

\begin{abstract}
We consider retarded settings in the context of a Bell-type experiment.  The retarded setting is defined as the value the setting would have taken were it not for some external intervention (for example, by a human). We derive \emph{retarded Bell inequalities} that explicitly take into account the retarded settings. These inequalities are not violated by Quantum Theory (or any other theory) when the retarded settings are equal to the actual settings.  We construct a simple model that reproduces Quantum Theory when the retarded and actual settings are equal, but violates it when they are not.  We discuss using humans to choose the settings in this type of experiment and the implications of a violation of Quantum Theory (in agreement with the retarded Bell inequalities) in this context.
\end{abstract}

\section{Introduction}

I first got interested in Bell's theorem \cite{bell1964einstein}, many years ago, on account of the following question: if we employed humans to switch the measurement settings at the two ends of the experiment, might we then expect Bell's inequalities to be satisfied and Quantum Theory to be violated?  I was particularly interested in whether we might think of this as a test for mind-matter duality.  The papers I wrote on this subject did not, of course, get past the referees in 1989.  In the meantime, I have come to be much more accepting of Bell style nonlocality in Quantum Theory. By now I more-or-less fully expect that, even if humans were used to switch the measurement settings, we would see a violation of Bell's inequalities in agreement with Quantum Theory.  On the other hand, the implications of a violation of Quantum Theory in this context would be so incredibly significant that it is worth discussing how we might go about doing an experiment.

In this contribution I will present modified Bell inequalities that I obtained 1989 (but did not publish) that take account of the possibility that a signal actually passes between the two ends of the experiment at the speed of light carrying information as to the distant setting (this is the \emph{retarded setting}). The inequalities I will present actually include these retarded settings.  After all, these are things we can measure and their values would be significant if switching distant settings actually changed the physics.

Although my original motivation for thinking about retarded settings was in the context of having people actually do the switching, we could use the inequalities obtained here in other contexts. For example, we might attempt to collect cosmological signals from regions of space-time that are causally disconnected from our own to implement the switching (see, for example, \cite{gallicchio2014testing}).  We might use them to analyze existing experiments in which the settings are varied in time \cite{aspect1982experimental} or in which a random number generator is used to do the switching \cite{weihs1998violation, scheidl2010violation}. Additionally, we might investigate practical applications of such inequalities (in device independent quantum cryptography \cite{barrett2005no, acin2006bell} and communication complexity \cite{cleve1997substituting, brukner2004bell} for example).

\section{Retarded settings}\label{sec:retardedsettings}

We will define two notions of retarded settings.  We are particularly interested in the second type (whose definition is a little subtle).  Consider that the settings $a$ and $b$ at the two ends, 1 and 2 respectively, of a Bell experiment are switched by some means during the course of the experiment.  Then we are interested in the retarded settings, $a_r$ (as regarded from side 2), and $b_r$ (as regarded from side 1).
\begin{description}
\item[Simple retarded settings.] The most obvious interpretation of retarded setting is that it is simply $a_r=a(t_2-L/c)$ where $L$ is the distance between the two ends and $t_2$ is the time at which the measurement at side $2$ takes place.  Likewise, we would have $b_r=b(t_1-L/c)$.
\item[Predictive retarded settings.]   A different notion of retarded settings is that $a_r$ is a prediction made at end 2 as to what setting $a$ will take at time $t_1$ based on information that can be locally communicated to end 2.  Thus, if the variation of $a$ were deterministic then a calculation at end 2 would enable us to predict $a$ at time $t_1$.  Now, we can imagine that the variation of $a$ is deterministic except for \emph{interventions}.  Then the retarded setting, $a_r$, is be the value $a$ is predicted to take at time $t_1$ if there are no interventions on this setting after time $t_2-L/c$ that alter $a$ at $t_1$ from the value it would have taken.
\end{description}
If we only allow the measurement setting to be changed by these supposed interventions then the above two definitions of retarded setting coincide. However, this will not be the case in general.

In the introduction we supposed that the interventions are due to a person doing the switching.  We will discuss this possibility later and the issues arising.  Another possibility, also mentioned above, is that the interventions are due to signals from causally disconnected regions of space.  One other possibility is that appropriate random number generators can supply such interventions.

\section{Clauser Horne Shimony Holt inequalities with retarded settings}

Consider a Bell type experiment with two ends.  Imagine we have a cental source of two systems, $1$ and $2$, described by hidden variables, $\lambda\in \Gamma$, with probability distribution $\rho(\lambda)$ such that
\begin{equation}
\int_\Gamma \rho(\lambda) d\lambda = 1
\end{equation}
We can obtain Clauser Horne Shimony Holt type Bell inequalities \cite{clauser1969proposed} with retarded settings.   In this scenario, we have a measurement, $A$, on system 1 which can take values $+1$ and $-1$.  Similarly, we have a measurement, $B$, on the right which can take values $+1$ and $-1$.  For simplicity, we will assume that the hidden variable model is deterministic (this assumption could easily be dropped).  Let
\begin{equation}  A(a, b_r, \lambda) \end{equation}
be the outcome at side 1 when we have setting $a$, retarded setting $b_r$, and hidden variable $\lambda$. Similarly,  we have
\begin{equation}
B(b, a_r, \lambda)
\end{equation}
at end 2. Note that, at each end, we allow for a dependence on the retarded setting at the other end. Since these settings are retarded, this is a local dependence.

We define the correlation function
\begin{equation}
E(a,b|a_r, b_r) = \int_\Gamma  A(a, b_r, \lambda) B(b, a_r, \lambda) d\lambda
\end{equation}
So $E(a,b|a_r, b_r)$ is the expectation value of the product of the outcomes at the two ends.

Clauser, Horne, Shimony, and Holt used the following (easily verified) mathematical identity
\begin{equation}\label{CHSHmathinequality}
 X'Y' +X'Y + XY' -XY =\pm 2
\end{equation}
where $X, X', Y, Y' =\pm 1$. We put
\begin{align}
X &= A(a, b_r, \lambda)\\
X'&=A(a', {b'}_r, \lambda)\\
Y &=B(b, a_r, \lambda) \\
Y'&=B(b', {a'}_r, \lambda)
\end{align}
Substituting these into \eqref{CHSHmathinequality} and integrating over $\lambda$ we obtain
\begin{equation}\label{CHSHretarded}
-2 \leq E(a',b'|{a'}_r, {b'}_r)+E(a',b|{a}_r, {b'}_r)+E(a,b'|{a'}_r, {b}_r)-E(a,b|a_r, b_r) \leq +2
\end{equation}
These are the retarded CHSH inequalities.

\section{When retarded and actual settings are equal}

If the retarded settings are equal to the actual settings for each term in the retarded CHSH inequalities \eqref{CHSHretarded} then we have ${a'}_r=a'=a_r=a$ and ${b'}_r=b'=b_r=b$ and the inequalities become
\begin{equation}
-2 \leq 2E(a,b|a,b) \leq +2
\end{equation}
This inequality is always satisfied (as $E$ is bounded by $\pm 1$) and hence there is no constraint from the retarded CHSH inequalities when the retarded settings are equal to the actual settings.

It is also interesting to consider the case where the retarded setting is equal to the actual setting for one end only.  Consider the case when ${a'}_r=a'=a_r=a$. Then the retarded CHSH inequality reduces to
\begin{equation}\label{CHSHsameoneend}
-2 \leq E(a,b'|a, {b'}_r)+E(a,b|a, {b'}_r)+E(a,b'|a, {b}_r)-E(a,b|a, b_r) \leq +2
\end{equation}
Now, this inequality is not violated by any theory, $T$, which has
\begin{equation}
E(a,b|a_r,b_r) = E^T(a,b)
\end{equation}
i.e.\ theories in which the retarded settings do not influence the physics (such as Quantum Theory).  This is because \eqref{CHSHsameoneend} then reduces to
\begin{equation}
-2 \leq 2 E^T(a,b')  \leq +2
\end{equation}
which cannot be violated.  Hence, if we want to test such theories, then we need to be sure the retarded and actual settings are different for each end.

\section{Testing Quantum Theory}

The quantum predictions do not depend on the retarded settings so, according to Quantum Theory, we would have
\begin{equation}\label{QuantPredE}
E(a,b|a_r, b_r) = E^\text{QT}(a,b)
\end{equation}
and the inequalities would become
\begin{equation}
-2 \leq E^\text{QT}(a',b')+E^\text{QT}(a',b)+E^\text{QT}(a,b')-E^\text{QT}(a,b) \leq +2
\end{equation}
As is well known, these inequalities can be violated by the predictions of Quantum Theory.

Hence, we cannot have a local model of the sort used in setting up the retarded CHSH inequalities that reproduces Quantum Theory.  However, we have also seen that we can have
\begin{equation}\label{property}
E(a,b|a,b) = E^\text{QT}(a,b)
\end{equation}
Furthermore, we would expect this to be true since the motivation for considering local models with retarded settings is to reproduce Quantum Theory when the retarded settings are equal to the actual settings.  The model we will provide in Sec.\ \ref{sec:amodel} has the property \eqref{property} by construction.   The retarded CHSH inequalities also allow
\begin{equation}
E(a,b|a,b_r) = E(a,b|a_r,b) = E^\text{QT}(a,b)
\end{equation}
(where the retarded setting equals the actual setting for one end).  The model we provide in Sec.\ \ref{sec:amodel} does not have this property. However, it should be possible to build a more sophisticated model that does have this property.

\section{Retarded versus standard Bell inequalities}

The standard Bell inequalities do not take account of retarded settings.  If we perform an experiment where we actively change the settings during the flight of the systems from the source before they arrive at the measurement apparatuses, then we have to take care to be sure that the retarded and actual settings are different for a sufficiently large proportion of cases. However, in the standard Bell inequalities, we simply ignore the retarded settings and average over all cases.   If the probability of any particular retarded setting is independent of the actual settings then we can recover standard Bell inequalities. In this case we can define
\begin{equation}
E_\text{av}(a,b)= \sum_{a_r,b_r}  p(a_r,b_r) E(a,b|a_r, b_r)
\end{equation}
where $p(a_r,b_r)$ is the probability that the retarded settings are $a_r$ and $b_r$.  Now we can take the average of the retarded CHSH inequality and obtain
\begin{equation}
-2 \leq E_\text{av}(a',b')+E_\text{av}(a',b)+E_\text{av}(a,b')-E_\text{av}(a,b) \leq +2
\end{equation}
These are standard CHSH inequalities (where we ignore the retarded settings).  However, this derivation of standard from retarded CHSH inequalities fails when there is a correlation between the retarded and actual settings.  Any such correlation could, in principle, lead to a situation where the standard Bell inequalities are violated while the retarded Bell inequalities are satisfied.  Hence, if we take seriously the need to actively switch the settings, then we need to use the retarded Bell inequalities.

It is particularly noteworthy that the famous experiment of Aspect, Dalibard and Roger in 1982 \cite{aspect1982experimental} used periodic switching.  Unfortunately, the switching period was such that the actual and retarded settings were equal.  This was pointed out by Zeilinger \cite{zeilinger1986testing} and formed part of the motivation for the experiment in his group \cite{weihs1998violation} in which ultrafast random switching was used.  Under such a scenario, it seems likely that the retarded and actual settings would not be correlated and hence we can obtain standard from retarded Bell inequalities by the above type of averaging (another, even more definitive, experiment was performed by Zeilinger's group in \cite{scheidl2010violation}).  On the other hand, if we use the retarded Bell inequalities directly, then we do not have to make such an assumption. Retarded Bell inequalities provide a tool for analyzing this kind of experiment.  Of course, neither of these experiments used humans or signals from cosmologically disconnected parts of the universe and so it is really models with \lq\lq simple retarded settings\rq\rq~ (as defined in Sec.\ \ref{sec:retardedsettings}) that are being tested (though one might argue that a random number generator forces interventions of the sort we discussed above.

Note added: very recently, an extraordinary experiment has been performed over 1.3km on the campus of Deft University of Technology that has switching and closes the detector efficiency loophole \cite{hensen2015experimental}.  A quantum random number generator is used to implement the switching.

\section{Source distribution of hidden variables}

In our model, we supposed that the retarded settings influenced the outcome at the other end (for example, by using a function $A(a, b_r, \lambda)$).  Another possibility (considered by Zeilinger in \cite{zeilinger1986testing}) is that the retarded settings influence the distribution of hidden variables at the source. Then we would have $\Gamma_{a_rb_r}$.  This would block the derivation of the retarded Bell inequalities above.  We can address this concern in the following way. First, rather than associating the hidden variables, $\lambda$, with the source alone we associate them with the full situation concerning the experiment at a time, $t_0$, earlier than both $t_1-L/c$ and $t_2-L/c$.  Thus, the hidden variables describe the source,  measurement apparatuses and every other detail of the physics that might be relevant for the experiment.  This means, in particular, that $\lambda$ also encodes the retarded settings $a_r$ and $b_r$ as long as there is no intervention between $t_0$ and the relevant retarded time.   Let us assume that these retarded settings are equal to $a$ and $b$.  If there are no interventions in the remaining time then the actual settings will be $a$ and $b$ respectively.  On the other hand, if there is an intervention at both ends, or just one end then we could have actual settings $a'$ and/or $b'$ accordingly.  This is true with the given initial distribution on $\lambda$ and so we can obtain retarded Bell inequalities as follows
\begin{equation}\label{sameretardedCHSH}
-2 \leq E(a',b'|a,b) + E(a',b|a,b) + E(a,b'|a,b) - E(a,b|a,b) \leq +2
\end{equation}
Note that every term has the same retarded settings (consistent with assumption above).  This inequality is interesting as only the $E(a',b'|a,b)$ term has different retarded and actual settings on both sides. The inequality can be violated by Quantum Theory if we substitute \eqref{QuantPredE} in.  Thus, in the unlikely event we saw a violation of Quantum Theory, the term, $E(a',b'|a,b)$, is the most likely to be the place we would see it.

\section{A model}\label{sec:amodel}

It is interesting to construct an explicit model reproducing the predictions of Quantum Theory (for a certain state) when the retarded settings are equal to the actual settings for both ends.  Consider a singlet state
\begin{equation}
|\psi\rangle=\frac{1}{\sqrt{2}} \left( |+\rangle_1|-\rangle_2 - |-\rangle_1|+\rangle_2 \right)
\end{equation}
We can subject this to a measurement of spin in the $xy$ plane at angle $a$ at end 1 and angle $b$ at end 2.  Then a simple calculation shows that the correlation function is
\begin{equation}
E^\psi(a, b) = - \cos(a-b)
\end{equation}
Now consider a hidden variable model with a hidden variable $\lambda$ having
\begin{equation}
0\leq \lambda < 2\pi ~~~~~~~ \Gamma= \frac{1}{2\pi}
\end{equation}
We define the result functions
\begin{equation}
A(a, b_r, \lambda) = \left\{ \begin{array}{ll}
                             +1 & \text{for}~ \theta_L \leq \lambda < \theta_L +\pi \\
                             -1 & \text{for}~ \theta_L \leq \lambda < \theta_L+2\pi
                             \end{array}  \right\}
\end{equation}
and
\begin{equation}
B(b, a_r, \lambda) = \left\{ \begin{array}{ll}
                             +1 & \text{for}~ \theta_R \leq \lambda < \theta_R +\pi \\
                             -1 & \text{for}~ \theta_R \leq \lambda < \theta_R+2\pi
                             \end{array}  \right\}
\end{equation}
where we understand $\lambda$ to be an angle (so angles greater than, or equal to $2\pi$ are identified with angles in the interval $[0, 2\pi)$ in the usual way) and where $\theta_L$ is a function of $a$ and $b_r$ and $\theta_R$ is a function of $b$ and $a_r$.  It is easy to prove that
\begin{equation}
E(a, b|a_r,b_r) = 1-\frac{2|\theta_R-\theta_L|}{\pi}
\end{equation}
Hence, if we set
\begin{equation}
\theta_L= -\frac{\pi}{4}(1+\cos(a-b_r))   ~~~~ \theta_R = \frac{\pi}{4}(1+\cos(a_r-b))
\end{equation}
we obtain
\begin{equation}\label{modeleqn}
E(a,b|a_r,b_r) = -\frac{1}{2}(\cos(a-b_r) + \cos(a_r-b))
\end{equation}
When the retarded settings are equal to the actual settings we get
\begin{equation}
E(a,b|a,b)=-\cos(a-b)
\end{equation}
in agreement with Quantum Theory.  If the actual and retarded setting differ for one side only then the model does not give quantum theory (although the retarded CHSH inequalities would allow quantum theory to be reproduced).

The retarded Bell inequalities are not violated by this model.  To illustrate this consider the special case
\begin{equation}
a= \frac{\pi}{2}, ~~ a' = 0, ~~ b=-\frac{\pi}{4}, ~~ b' = \frac{\pi}{4}
\end{equation}
If we substitute
\eqref{modeleqn} into \eqref{sameretardedCHSH} with these settings then we obtain
\begin{equation}
E(a',b'|a,b) + E(a',b|a,b) + E(a,b'|a,b) - E(a,b|a,b) = -\sqrt{2}
\end{equation}
This satisfies the particular retarded CHSH inequalities.  It is interesting that we do not saturate the inequalities with this model.  A better model may saturate the inequality.

\section{Clauser Horne inequalities}

We can also derive retarded Clauser Horne inequalities based on the Clauser Horne inequalities \cite{clauser1974experimental}.  These inequalities are especially useful in experiments since they have $0$ as the upper bound. Consequently it is sufficient to measure count rates without normalizing the probabilities with a total count rate.  These inequalities pertain to the same setting as before, but now we are interested in the probabilities for some particular outcome (we will take this to be the $+$ outcome) at each end.  We let
\begin{equation}
p_1(a, b_r|\lambda)
\end{equation}
be the probability of that we see a outcome $+1$ to measurement $A$ with setting $a$ at this end and retarded setting $b_r$ at the other end.   Similarly we have
\begin{equation}
p_2(b, a_r|\lambda)
\end{equation}
for the probability that we see outcome $+1$ for measurement $B$ on particle 2 with setting $b$ and retarded setting $a_r$ at the other end.   The joint probability of seeing a $+1$ at both ends is
\begin{equation}
p_{12}(a, b|{a}_r, {b}_r) = \int_\Gamma p_1(a, b_r|\lambda) p_2(b, a_r|\lambda) d\lambda
\end{equation}
Note that we allow for a dependence of this joint probability on the retarded settings at the other end.   We can also construct the local probabilities
\begin{align}
p_1(a) &= \int_\Gamma p_1(a, b_r|\lambda) d\lambda \\
p_2(b) &= \int_\Gamma p_2(b, a_r|\lambda )d\lambda
\end{align}
We could, without violating locality, also allow these probabilities to depend on the retarded settings.  However, this seems less likely and so we will stick with the given functional dependence.  If we do want to have such functional dependence then this can easily be inserted in the Bell inequalities we derive below.

Quantum theory does not predict any dependence on the values of retarded settings. Thus, according to Quantum Theory, we will have
\begin{equation}\label{QuantPredprob}
p_{12}(a, b|{a}_r, {b}_r) = p^\text{QT}_{12}(a, b)
\end{equation}
However, it will follow (by adapting the usual Bell analysis) that this cannot actually be the case in a local hidden variable model of the type we are considering.

Now we will derive retarded Clauser Horne inequalities.  Consider the following easily verified mathematical inequalities (introduced by Clauser and Horne \cite{clauser1974experimental})
\begin{equation}\label{CHxyinequalities}
-1 \leq x'y' + x'y +xy' -xy -x-y \leq 0
\end{equation}
where $0\leq x,y,x',y' \leq 1$.  Now we put
\begin{align*}
x &= p_1(a, b_r|\lambda) \\
x'&=p_1(a', {b'}_r|\lambda) \\
y &=p_2(b, a_r|\lambda)  \\
y'&=p_2(b, {a'}a_r|\lambda)
\end{align*}
Inserting these into \eqref{CHxyinequalities} we obtain
\begin{align}\label{CHretinequalities}
-1 \leq p_{12}(a', b'|{a'}_r, {b'}_r) + p_{12}(a', b|{a}_r, {b'}_r) +  p_{12}(a, b'|{a'}_r, {b}_r)& \nonumber \\
 - p_{12}(a, b|{a}_r, {b}_r) - p_{1}(a)& - p_2(b) \leq 0
\end{align}
Note that we could allow $p_1(a)$ and $p_2(b)$ to depend on the retarded settings (so we would have $p_1(a|b_r)$ and $p_2(b|a_r)$ instead).   We can substitute the quantum predictions \eqref{QuantPredprob} in to this inequality. This gives
\begin{equation}
-1 \leq p_{12}^\text{QT}(a', b') + p_{12}^\text{QT}(a', b) + p_{12}(a, b') - p_{12}^\text{QT}(a, b) - p_{1}^\text{QT}(a) - p_2^\text{QT}(b) \leq 0
\end{equation}
It was shown by Clauser and Horne that this inequality can be violated by choosing the two systems to be in an appropriate entangled state and by choosing appropriate settings. Note that, to get a violation by Quantum Theory, it is a necessary condition that both $a'\not=a$ and $b'\not=b$.

\section{How to perform an experiment}

To perform an experiment to test the retarded Bell inequalities we need a source of interventions.  We could imagine two subjects, let us call them Alice and Bob, sitting at the two ends each switching the settings by hand.  The only problem with this is that the hand and the device it switches both operate at mechanical speeds.  To have any chance of having a retarded setting different from the actual setting, with such a system, we would need the distance between the two ends to be very big.  However, whenever Alice decides to switch the setting, there is some accompanying electrical activity in the brain.  We could use this accompanying electrical activity to do the actual switching (where the device that Alice switches is just a retrospective control).  Pockel cells can be used to accomplish fast switching (at electrical speeds) on photons. Hence, with this set up we could realistically perform an experiment over a shorter distance.

It is not actually necessary that Alice (and Bob) actually switch a switch. It only necessary that they engage in some activity such that we want to regard the resulting electrical activity as constituting an intervention.  A challenge would be identifying the electrical signals originating in the brain that should be regarded as interventions in the sense understood here.  However, we could filter for lots of different types of signature and analyze the data accordingly.  Another challenge would be to get the count rates high enough that we can have statistically meaningful results.

A possible control on this type of experiment would be to introduce a delay between the source of interventions and the switching.  If this delay were longer than $L/c$ then any supposed effect ought to vanish.

\section{Interpretation}

Quantum Theory is a coherent whole.  It has been tested extensively at low energy in laboratories around the world.  Hence, it seems very unlikely that we could expect to see its violation in circumstances such as those we have discussed. On the other hand, as long as we do these kind of experiments to test the theory, it is worth thinking carefully about what we are testing.  We have here suggested that we may see a violation of Quantum Theory in accordance with the retarded Bell inequalities derived here when we actually use humans to switch the settings.  If such a violation of Quantum Theory was seen, and yet it was impossible to obtain such a violation where the switching was performed by non-animate systems (such as computer programs, physically chaotic systems, or quantum random number generators) then we would have to seek an explanation of this.  The Cartesian idea of mind-matter duality provides a model for a kind of external intervention (of mind on matter) of the type that we have discussed.  This kind of duality has been much discussed by philosophers in the context of understanding consciousness.  A modern proponent of such dualism is Chalmers \cite{chalmers1997conscious} while Dennett \cite{dennett1993consciousness} advocates the opposite point of view.   While it is difficult to understand consciousness in terms of matter stuff alone, it is not clear that adding mind stuff into the mix helps us particularly. On the other hand, if it turned out that systems we take to be conscious were capable of things (like violating Quantum Theory under the described circumstances) that ordinary systems were not then that would be a profound challenge to our usual way of thinking about the world.  We should not shy away from such experiments.

The situation here is reminiscent of the Turing test \cite{turing1950computing}.  In the Turing test, computers and humans compete over a computer screen interface to convince human interviewers that they are human.  This test involves the subjective judgement of the interviewers.   Here, instead, pairs of humans compete against pairs of computers (or whatever other physical system we want to use) to violate Quantum Theory by providing the inputs to the setting switchs of an experimental apparatus which they, otherwise, have no control over.  The test is completely objective - passing the test would entail bringing about a violation of Quantum Theory in accord with the retarded Bell inequalities.  As such, this test is interesting simply because it provides a scientific way to investigate a particular model of mind-matter duality (even if, as seems overridingly likely, an actual test will not violate Quantum Theory).

\section{Bell's La Nouvelle Cuisine remarks}

I never met John Bell (though I was in the audience for two talks he gave).  I did, however, send him a copy of a paper outlining some of the above ideas (my second attempt attempt at such a paper).  He responded by sending me a copy of his La Nouvelle Cuisine paper \cite{sarlemijn2012nouvelle} (now available as the penultimate article in the wonderful collection of papers by Bell in \cite{bell2001john}).  This paper is a beautiful discussion of how to understand causality.  While he had clearly thought about using humans to do the switching, I suspect he was not sympathetic to the idea that anything would come of it.  I end with a quote from this paper - it is classic Bell:
\begin{quote}
The assertion that \lq\lq we cannot signal faster than light\rq\rq~ immediately provokes the question:
\begin{quote}
Who do we think \emph{we} are?"
\end{quote}
\emph{We} who can make \lq\lq measurements\rq\rq, \emph{we} who can manipulate \lq\lq external fields\rq\rq, \emph{we} who can \lq\lq signal\rq\rq~ at all, even if not faster than light? Do \emph{we} include chemists, or only physicists, plants, or only animals, pocket calculators, or only mainframe computers?

The unlikelihood of finding a sharp answer to this question reminds me of the relationship of thermodynamics to fundamental theory.
\end{quote}

\section*{Acknowledgements}

I am grateful to the late Euan Squires for discussions on this subject while I was doing my PhD under his supervision. Research at Perimeter Institute is supported by the Government of Canada through Industry Canada and by the Province of Ontario through the Ministry of Economic Development and Innovation. This project was made possible in
part through the support of a grant from the John Templeton Foundation. The opinions expressed in this publication are those of the author and do not necessarily
reflect the views of the John Templeton Foundation.

\bibliography{BellBib}
\bibliographystyle{plain}

\end{document}